# Quantum secure dialogue with quantum encryption

Tian-Yu Ye*

College of Information & Electronic Engineering, Zhejiang Gongshang University, Hangzhou 310018, P.R.China

**Abstract**

How to solve the information leakage problem has become the research focus of quantum dialogue. In this paper, in order to overcome the information leakage problem in quantum dialogue, a novel approach for sharing the initial quantum state privately between communicators, i.e., quantum encryption sharing, is proposed by utilizing the idea of quantum encryption. The proposed protocol uses EPR pairs as the private quantum key to encrypt and decrypt the traveling photons, which can be repeatedly used after rotation. Due to quantum encryption sharing, the public announcement on the state of the initial quantum state is omitted, thus the information leakage problem is overcome. The information-theoretical efficiency of the proposed protocol is nearly 100%, much higher than previous information leakage resistant quantum dialogue protocols. Moreover, the proposed protocol only needs single-photon measurements and nearly uses single photons as quantum resource so that it is convenient to implement in practice.

**PACS:** 03.67.Dd; 03.65.Ud

**Keywords:** Quantum dialogue, quantum encryption, information leakage.

## 1 Introduction

Quantum cryptography has been an interesting branch of quantum information processing, which is based on the law of quantum mechanics. Until now, quantum cryptography has been greatly pursued so that a lot of good quantum cryptography protocols have been proposed by some groups from different viewpoints, including quantum key distribution (QKD)[1-3], quantum secret sharing (QSS)[4-6], quantum secure direct communication (QSDC)[7-14], *et al*. Recently, a special kind of quantum cryptography protocol emerged, i.e., the channel-encrypting (quantum-encrypting) quantum cryptography protocol[15-21]. In the literature of quantum-encrypting quantum cryptography, communicators always share a sequence of private quantum states, which play the role of their reusable quantum key, and use them to encrypt and decrypt the classical secret messages carried by the traveling states.

According to the communication mode, QSDC can always be divided into two kinds, i.e., one-way QSDC and two-way QSDC. All of the protocols in Refs.[7-14] belong to the first kind. The two-way QSDC is also called as quantum dialogue, independently put forward by Zhang *et al.*[22-23] and Nguyen[24] firstly, where communicators are able to accomplish the mutual exchange of their respective classical secret messages simultaneously. In some works of this literature [22-34], in order to accomplish the dialogue, two public announcements are always needed. That are, the announcement on the state of the initial quantum state, and the announcement on the state of the quantum state encoded with all communicators' secret messages[22,24-28,30,32,34] (or the announcement on the entanglement swapping outcomes between the initial quantum state and the quantum state encoded with all communicators' secret messages[23,29,31], or the announcement on the entanglement swapping outcomes between two quantum states individually encoded with one communicator' secret messages[33] ). However, once Eve hears of both public announcements, she may obtain partial information about all communicators' secret messages, as there always exists 'classical correlation' [35] between them. As a result, 'information leakage' [36-39] takes place, which means that partial of secret messages leak out to Eve without any active attacks. As the latter announcement is necessary for dialogue, the only way to prevent the information leakage problem lies in avoiding the first announcement, i.e., the one on the state of the initial quantum state. However, for successful dialogue, the communicators, who do not prepare the initial quantum state, need to know the state of the initial quantum state. Hence, the feasible solution to avoid the first announcement is to share the initial quantum state privately among all communicators. Fortunately, at present, some approaches for doing so in quantum dialogue have been suggested, such as the direct transmission of initial quantum state[40,41,44,45], the correlation extractability of Bell states[42] and the measurement correlation from entanglement swapping of two Bell states[43].

In this paper, a novel approach for sharing the initial quantum state privately between communicators, i.e., quantum encryption sharing, is proposed by utilizing the idea of quantum encryption. The proposed protocol uses EPR pairs as the private quantum key to encrypt and decrypt the traveling photons, which can be repeatedly used after rotation. Due to quantum encryption sharing, the public announcement on the state of the initial quantum state is omitted, thus the information leakage problem is overcome. The information-theoretical efficiency of the proposed protocol is nearly 100%, much higher than previous information leakage resistant quantum dialogue protocols. Moreover, the proposed protocol only needs single-photon measurements and nearly uses single photons as quantum resource so that it is convenient to implement in practice.

## 2 Quantum dialogue protocol

Suppose that Alice has $N$ bits secret messages $\{r_1, r_2, \cdots, r_N\}$, and Bob has $N$ bits secret messages $\{k_1, k_2, \cdots, k_N\}$,

---

*Corresponding author:
E-mail：happyyty@aliyun.com(T.Y.Ye)

where $r_i, k_i \in \{0,1\}$ $(i = 1, 2, \cdots, N)$. Alice and Bob agree on in advance that $U_0$ and $U_1$ denote two unitary operations $I = |0\rangle\langle 0| + |1\rangle\langle 1|$ and $\sigma_x = |0\rangle\langle 1| + |1\rangle\langle 0|$, respectively. The proposed quantum dialogue protocol consists of the following steps.

**Step1: Quantum key generation and distribution.** Alice prepares $N$ two-photon pairs $\{(A_1, B_1), (A_2, B_2), \cdots, (A_N, B_N)\}$ all in the state of EPR pair

$$|\Phi\rangle_{AB} = \frac{1}{\sqrt{2}}(|00\rangle + |11\rangle)_{AB}. \tag{1}$$

Alice extracts the photon $A$ in each pair to form the sequence $S_A$, hence $S_A = \{A_1, A_2, \cdots, A_N\}$. The other photons compose of the sequence $S_B$, i.e., $S_B = \{B_1, B_2, \cdots, B_N\}$. For checking eavesdropping, Alice adds a small trick in $S_B$. That is, she inserts some decoy photons into $S_B$, which are randomly in one of the four states $\{|0\rangle, |1\rangle, |+\rangle = \frac{1}{\sqrt{2}}(|0\rangle + |1\rangle), |-\rangle = \frac{1}{\sqrt{2}}(|0\rangle - |1\rangle)\}$. As a result, $S_B$ turns into $S_B'$. Then, Alice sends $S_B'$ to Bob and keeps $S_A$ by herself. After Bob announces the receipt of $S_B'$, Alice tells Bob the positions and the preparation bases of the decoy photons. Bob measures the decoy photons in the same bases as Alice told, and informs Alice of his measurement outcomes. Alice calculates the error rate by comparing the initial states of the decoy photons with Bob's measurement outcomes. If the error rate is low enough, they have securely shared the private quantum key and go on the next step; otherwise, they restart from the beginning.

**Step2: Alice's encryption.** Alice prepares a sequence of $N$ traveling photons $S_P = \{P_1, P_2, \cdots, P_N\}$ with states $\{|m_1\rangle, |m_2\rangle, \cdots, |m_N\rangle\}$, where $m_i = 0$ or $1$ $(i = 1, 2, \cdots, N)$. For checking eavesdropping, Alice also inserts some decoy photons into $S_P$, which are randomly in one of the four states $\{|0\rangle, |1\rangle, |+\rangle, |-\rangle\}$. As a result, $S_P$ turns into $S_P'$. Alice uses the private quantum key $|\Phi\rangle_{AB}$ to encrypt the traveling photons in $S_P'$ except for decoy photons. That is, Alice performs a CNOT operation on the photons $A_i$ and $P_i$ $(i = 1, 2, \cdots, N)$ by using the photon $A_i$ as the control qubit and $P_i$ as the target qubit. Consequently, the three photons $A_i$, $B_i$ and $P_i$ are in the GHZ state

$$|\Psi\rangle = \frac{1}{\sqrt{2}}(|00m_i\rangle + |11\bar{m}_i\rangle)_{A_i B_i P_i}, \tag{2}$$

where $\bar{m}_i = 1 - m_i$. Then, Alice sends $S_P'$ to Bob. After Bob announces the receipt of $S_P'$, Alice and Bob implement the security check procedure in the same way as they did in Step 1. After they confirm that the transmission of $S_P'$ is secure enough, they go on the next step.

**Step3: Bob's decryption and encoding.** Bob drops out the decoy photons in $S_P'$ firstly. As a result, $S_P'$ turns back into $S_P$. Then, Bob decrypts the traveling photons in $S_P$. That is, Bob performs a CNOT operation on the photons $B_i$ and $P_i$ by using the photon $B_i$ as the control qubit and $P_i$ as the target qubit. Consequently, the two photons $A_i$ and $B_i$ and the traveling photon $P_i$ are in the same state as the initial state

$$|\Psi'\rangle = |\Phi^+\rangle_{A_i B_i} \otimes |m_i\rangle_{P_i} \tag{3}$$

Then, Bob measures the traveling photon $P_i$ with $Z$-basis $(\{|0\rangle, |1\rangle\})$ to know the initial state of $P_i$. According to his measurement outcome, Bob reproduces a new $P_i$ with no measurement performed. Then, Bob performs the unitary operation $U_{k_i}$ on the new $P_i$ to encode his one-bit secret message $k_i$. Accordingly, $P_i$ turns into $U_{k_i} P_i$. For checking eavesdropping, Bob inserts some decoy photons into $S_P$, which are randomly in one of the four states $\{|0\rangle, |1\rangle, |+\rangle, |-\rangle\}$. As a result, $S_P$ turns into $S_P''$. Then, Bob sends $S_P''$ to Alice. After Alice announces the receipt of $S_P''$, Bob and Alice also implement the security check procedure in the same way as they did in Step 1. After they confirm that the transmission of $S_P''$ is secure enough, they go on the next step.

**Step4: Alice's encoding and quantum dialogue.** Alice drops out the decoy photons in $S_P''$ firstly. As a result, $S_P''$ turns back into $S_P$. Then, Alice performs the unitary operation $U_{r_i}$ on $U_{k_i} P_i$ in $S_P$ to encode her one-bit secret message $r_i$. Accordingly, $U_{k_i} P_i$ turns into $U_{r_i}U_{k_i}P_i$. Afterward, Alice measures $U_{r_i}U_{k_i}P_i$ with $Z$-basis. For dialogue, Alice publicly announces the measurement outcome of $U_{r_i}U_{k_i}P_i$. According to the initial state of $P_i$, her own unitary operation $U_{r_i}$ and the measurement outcome of $U_{r_i}U_{k_i}P_i$, Alice can read out Bob's one-bit secret message $k_i$. Likewise, Bob can also read out Alice's one-bit secret message $r_i$.

**Step5: Quantum key rotation.** For each of their shared EPR pairs $|\Phi\rangle_{AB}$, Alice and Bob rotate their photon's state by angle $\theta$, respectively. The rotation can be described as

$$R(\theta) = \begin{pmatrix} \cos\theta & \sin\theta \\ -\sin\theta & \cos\theta \end{pmatrix} \qquad (4)$$

Although the bilateral operation of $R(\theta)$ does not change the state of $|\Phi\rangle_{AB}$, it can prevent the other party from eavesdropping [15] (The selection of $\theta$ will be given later in the security analysis section). After the rotation, Alice and Bob can use the EPR pairs $|\Phi\rangle_{AB}$ repeatedly as their private quantum key in the next round, and repeat their procedures from Step 2.

Suppose that the first bits of Alice's and Bob's secret messages are **0** and **1**, respectively. In other words, $r_1 = 0$ and $k_1 = 1$. Moreover, suppose that Alice prepares the first traveling photon $P_1$ in the state of $|0\rangle$. That is, $m_1 = 0$. Alice prepares the quantum key $|\Phi\rangle_{A_1B_1}$ and shares it with Bob. Then, Alice uses the photon $A_1$ to encrypt $P_1$ through the CNOT operation. Afterward, Bob uses the other photon $B_1$ to decrypt out $P_1$ through the CNOT operation. Then, Bob measures $P_1$ with $Z$-basis, thus he can know the initial state of $P_1$. According to his measurement outcome, Bob reproduces a new $P_1$ with no measurement performed. As a result, the new $P_1$ evolves as follows:

$$|0\rangle_{P_1} \Rightarrow \sigma_x \otimes |0\rangle_{P_1} = |1\rangle_{P_1} \Rightarrow I \otimes |1\rangle_{P_1} = |1\rangle_{P_1} \qquad (5)$$

According to three known information: the announced measurement outcome $|1\rangle_{P_1}$, the prepared initial state $|0\rangle_{P_1}$ and her unitary operation $I$, Alice can read out that $k_1$ is **1**. Similarly, Bob can read out that $r_1$ is **0**, according to the announced measurement outcome $|1\rangle_{P_1}$, the prepared initial state $|0\rangle_{P_1}$ and his unitary operation $\sigma_x$.

### 3 Security analysis

In this section, the security is analyzed with regard to different steps of this protocol.

**Quantum key generation and distribution.** In this step, Alice prepares EPR pairs in the state of $|\Phi\rangle_{AB}$ as the private quantum key, and sends the photons $B$ to Bob. Eve may intend to obtain the photons $B$ during this transmission, and use them to decrypt the ciphertext Alice sends to Bob subsequently. If Eve can succeed in doing it, she will know the initial state of $P_i$ so that she can legally obtain partial information about Alice and Bob's secret messages after Alice's announcement on the measurement outcome of $U_{r_i}U_{k_i}P_i$. Fortunately, Eve is not able to accomplish her aim. The reason lies in the following two aspects:[21] (1) the quantum no-cloning theorem guarantees that Eve has no chance to replicate the photons $B$; (2) Eve can not distinguish between the decoy photons and the photons $B$, since they are in the same state, i.e., the maximally mixed state $\rho = 1/2(|0\rangle\langle 0| + |1\rangle\langle 1|)$. Consequently, although Eve expects to impose her attacks only on the photons $B$, her attacks will be inevitably performed on decoy photons. Hence, any attack from Eve will leave a trace on the decoy photons and be detected by Alice and Bob. That is to say, the distribution process of quantum key is secure in principle. Note that the idea of using decoy photons for eavesdropping check, which is derived from the BB84 protocol [1], has been widely used in some previous protocols [19,20,21,27,28,31,32,34,40,44,45], and its validity has been verified.

**Alice's encryption.** In this step, Alice encrypts the traveling photons and sends the ciphertext to Bob. The encryption on a traveling photon $P_i$ with the CNOT operation by Alice makes it entangle with a quantum system in the quantum key $|\Phi^+\rangle_{A_iB_i}$. After the CNOT operation done by Alice, the three photons $A_i$, $B_i$ and $P_i$ are in the GHZ state $|\Psi\rangle = \frac{1}{\sqrt{2}}(|000\rangle + |111\rangle)_{A_iB_iP_i}$ or $|\Psi\rangle = \frac{1}{\sqrt{2}}(|001\rangle + |110\rangle)_{A_iB_iP_i}$ randomly. Apparently, the reduced density matrix of the traveling photons $P_i$ is $\rho_{P_i} = 1/2\begin{pmatrix} 1 & 0 \\ 0 & 1 \end{pmatrix}$. That is to say, the traveling photons $P_i$ in $S_P$ are always in the maximal mixture of $|0\rangle_{P_i}$ and $|1\rangle_{P_i}$ during the transmission. Therefore, Eve cannot elicit any useful information just from the ciphertext even if she intercepts it. In addition, if Eve wants to decrypt out the initial state of $P_i$, she needs to do one more thing, i.e., eavesdropping the photon $B_i$ during the distribution process of quantum key. However, as analyzed above, the quantum no-cloning theorem and the usage of decoy photons ensure that Eve cannot accomplish this aim.

In fact, Eve cannot intercept the ciphertext without being discovered. The reason also lies in the quantum no-cloning theorem and the usage of decoy photons in this step.

**Bob's decryption and encoding.** In this step, Bob decrypts out the initial states of the traveling photons, encodes them with his secret messages and sends the encoded traveling photons to Alice. Eve cannot elicit Bob's one-bit secret message from the encoded traveling photon $U_{k_i}P_i$ even if she intercepts it, since she has no knowledge about the initial state of $P_i$. In fact, Eve cannot intercept the encoded traveling photon $U_{k_i}P_i$ without being discovered, due to the quantum no-cloning theorem and the usage of decoy photons in this step.

**Alice's encoding and quantum dialogue.** In this step, Alice conducts the encoding of her secret messages, and accomplishes the dialogue with Bob. Apparently, no photons are transmitted in this step so that Eve has no chance to attack.

**Quantum key rotation.** In this step, both Alice and Bob rotate their photons' state. In Ref.[15], a QKD with quantum encryption is proposed, where Eve may take the following attack strategy. That is, Eve firstly entangles her ancilla into the quantum key $|\Phi\rangle_{AB}$ shared by Alice and Bob by intercepting the traveling particle $\gamma$ Alice sends to Bob and performing a CNOT operation on her ancilla with $\gamma$ being the control qubit, and then utilizes this entangled relation to elicit useful information about the key bits. In Ref.[16], Gao et al presented another special attack strategy for Eve to attack the QKD protocol of Ref.[15], which also threatens other QKD protocol based on reused quantum key, such as the one of Re.[17]. In addition, it has also been proven in Ref.[16] that, as long as $\theta \neq \theta_1$ ($\theta_1 = k\pi \pm \pi/4, k = 0, \pm 1, \pm 2, \cdots$), it is inevitable for Eve to introduce disturbance when she takes either the attack strategy of Ref.[15] or the special attack strategy of Ref.[16]. Therefore, in the proposed protocol, before each new round, Alice and Bob need to rotate their photon's state by choosing a proper $\theta$, respectively, for each of their shared EPR pairs $|\Phi\rangle_{AB}$. In this way, the EPR pairs $|\Phi\rangle_{AB}$ can be repeatedly used as their private quantum key in the following runs.

## 4 Discussions
**(1) The information leakage problem**

Without loss of generality, the author also takes the case in section 2 as example here. With the help of private quantum key $|\Phi\rangle_{A_1B_1}$, Bob can know the initial state of $P_1$, making Alice unnecessary to publicly announce the initial state of $P_1$. As a result, Eve is unsure about the initial state of $P_1$ so that she has to guess it when she hears from the measurement outcome of $|1\rangle_{P_1}$. If she guesses that the initial state is $|0\rangle_{P_1}$, the secret messages are $\{r_1=0, k_1=1\}$ or $\{r_1=1, k_1=0\}$; if she guesses that the initial state is $|1\rangle_{P_1}$, the secret messages are $\{r_1=0, k_1=0\}$ or $\{r_1=1, k_1=1\}$. Apparently, there are four kinds of uncertainty, corresponding to $-\sum_{i=1}^{4} p_i \log_2 p_i = -4 \times \frac{1}{4} \log_2 \frac{1}{4} = 2$ bit information for Eve. It means that none of the two bits secret messages $\{r_1, k_1\}$ have been leaked out. Therefore, the proposed protocol has successfully avoided the information leakage problem.

**(2) The information-theoretical efficiency**

The information-theoretical efficiency[3] is defined as $\eta = b_s/(q_t + b_t)$, where $b_s$, $q_t$ and $b_t$ are the expected secret bits received, the qubits used and the classical bits exchanged between Alice and Bob. In the proposed protocol, the first traveling photon $P_1$ can be used for exchanging Alice and Bob's respective one bit secret message with one classical bit needed for the announcement on the measurement outcome of $U_{r_1}U_{k_1}P_1$ and two qubits needed for the private quantum key $|\Phi\rangle_{A_1B_1}$. Since the private quantum key $|\Phi\rangle_{A_1B_1}$ can always be repeatedly used in the following runs, the two qubits it consumes can almost be ignored. Hence, the information-theoretical efficiency of the proposed protocol is extremely near to $100\%$.

**(3) Comparisons of previous information leakage resistant quantum dialogue protocols**

Since the proposed protocol has no information leakage problem, its comparisons with previous information leakage resistant quantum dialogue protocols [40-45] are made here. The comparisons are summarized in Table 1, with respect to the initial quantum resource, the quantum measurement and the information-theoretical efficiency. ①The initial quantum resource. The proposed protocol consumes single photons as the traveling photons and Bell states as the private quantum key. As the private quantum key can be repeatedly used, the Bell states consumed in the proposed protocol may be ignored. Moreover, single photons are much easier to prepare than Bell states and GHZ states. Hence, with respect to the initial quantum resource, according to Table 1, the proposed protocol exceeds all protocols of Refs.[40,42-45], and is nearly the same as the protocol of Ref.[41]. ②The quantum measurement. Obviously, the proposed protocol only needs the single-photon measurements. As the single-photon measurements are much easier to perform in practice than both the Bell-basis measurements and the GHZ-basis measurements, with respect to the quantum measurement, according to Table 1, the proposed protocol exceeds all protocols of Refs.[40,42-45], and has the same performance as the protocol of Ref.[41]. ③The information-theoretical efficiency. According to Table 1, the proposed protocol has the highest information-theoretical efficiency, since its efficiency is nearly 100%.

From the above analysis, it can be concluded that compared with those protocols in Refs.[40-45], from an overall view, the proposed protocol takes advantage over them in the initial quantum resource, the quantum measurement and the information-theoretical efficiency.

Table 1. Comparisons of previous information leakage resistant quantum dialogue protocols

|  | The protocol of Ref.[40] | The protocol of Ref.[41] | The protocol of Ref.[42] | The protocol of Ref.[43] | The protocol of Ref.[44] | The protocol of Ref.[45] | The proposed protocol |
|---|---|---|---|---|---|---|---|
| Initial quantum resource | Bell states | single photons | Bell states and single photons | Bell states | GHZ states | Bell states | Nearly single photons |
| Quantum measurement | Bell-basis measurements | single-photon measurements | single-photon measurements and Bell-basis measurements | Bell-basis measurements | GHZ-basis measurements | Bell-basis measurements | single-photon measurements |
| Information-theoretical efficiency | 66.7% | 66.7% | 75% | 66.7% | 66.7% | 66.7% | Nearly 100% |

## 5 Conclusions

In all, the proposed protocol has several distinct features:

① It avoids the public announcement on the state of the initial quantum state by using a novel approach for sharing the initial quantum state privately between communicators, i.e., quantum encryption sharing. Thus the information leakage problem in quantum dialogue is overcome here.

② It can repeatedly use the private quantum key after doing rotation, thus the quantum resource is economized and the efficiency is enhanced.

③ Its information-theoretical efficiency is nearly 100%, much higher than previous information leakage resistant quantum dialogue protocols.

④ It only needs single-photon measurements and nearly uses single photons as quantum resource so that it is convenient to implement in practice.

**Acknowledgements**

Funding by the National Natural Science Foundation of China (Grant No.11375152), and the Natural Science Foundation of Zhejiang Province (Grant No.LQ12F02012) is gratefully acknowledged.